\documentstyle{l-aa}

\begin{document}

   \thesaurus{ A\&A Section 6 
             (06.16.1; 08.05.3; 08.07.1)} 

   \title{Stars as galactic neutrino sources}

 \author{E. Brocato \inst{1}$^,$ \inst{2}, V. Castellani \inst{1}$^,$ \inst{3},
 S. Degl'Innocenti \inst{3}$^,$\inst{4} G. Fiorentini \inst{4}$^,$\inst{5}, 
G. Raimondo \inst{1}$^,$ \inst{2}}

   \offprints {S. Degl'Innocenti, Dipartimento di Fisica Universit\`a 
di Pisa, piazza Torricelli 2, 56126 Pisa, Italy, scilla@astr1pi.difi.unipi.it}

\institute{
  Osservatorio Astronomico di Collurania, via Mentore Maggini I-64100 Teramo, Italy
\and Istituto Nazionale di Fisica Nucleare, LNGS, I-67100 L'Aquila, Italy  
\and Dipartimento di Fisica dell'Universit\'a di Pisa, Piazza Torricelli 2,
   I-56126 Pisa, Italy 
   \and Istituto Nazionale di Fisica Nucleare,
     Sezione di Ferrara, via Paradiso 12, I-44100 Ferrara, Italy
 \and Dipartimento di Fisica dell'Universit\`a di Ferrara,
     via Paradiso 12, I-44100 Ferrara, Italy
		}

   \date{Received ......... ; accepted ..........   }

   \maketitle

   \markboth {Brocato et al.: Stellar neutrinos}{Brocato et al.: Stellar Neutrinos}

   \begin{abstract}

Theoretical expectations concerning stars as neutrino 
sources are presented according to detailed evaluations of the stellar
evolutionary histories for an extended grid of 
stellar masses. Neutrino fluxes and cumulative neutrino yields
are given for both `thermo-nuclear' and `cooling' neutrinos all over the
nuclear life of the stars and along the final cooling as White Dwarfs.
Predictions concerning the galactic and the cosmic 
neutrino background are presented and discussed.


   \end{abstract}



The fusion of H into He in the
solar interior has to be accompanied by the emission of about
10$^{38}$ neutrinos per second. Thus at the earth surface one expects
a flux of the order of 10$^{11}$ neutrinos cm$^{-2}$ sec$^{-1}$.
The successful detection of these solar neutrinos in several experiments
over the world has raised  the attention to the Sun as a neutrino source,
originating in the same 
time (see e.g. Bahcall 1989, Bahcall et al. 1995) the long debated problem of the discrepancy
between experimental data and theoretical predictions.

All stars that are burning 
and/or have been burning H in the Universe can be regarded as
cosmic sources of both photons and neutrinos, thus simultaneously
contributing to the background of those particles. Clearly this stellar
 neutrino flux is expected to be much smaller 
than the solar one; however the amount and the energy of stellar neutrinos
contributing to the neutrino background appears as an interesting question
to be addressed in the frame of the present theoretical knowledge of stellar
evolutionary structures. 
A first step in such a direction has been recently presented by
Hartmann et al. (1994), assuming all stars in their initial (Main
Sequence, MS) phase. A quick inspection of the stellar
evolutionary scenario reveals that such an assumption has to be
regarded only as a first approximation to the problem. As a matter
of fact, we know, e.g.,that low mass main sequence stars burn
H into He through the proton-proton chain, thus emitting a mixture
of neutrinos from pp, $^7$Be and $^8$B reactions. In the meantime,
we  know that in the same structures the largest amount of He is 
produced at higher temperatures in the Red Giant stage, 
where H burning is dominated by CNO reactions. Accordingly, one can
easily predict that all along  the life of a low mass star CNO
neutrinos will eventually dominate the yield, in spite of the MS
behaviour.\\
Moreover, the burning of H into He is not the only way used by
 stars to produce neutrinos. In the advanced phases 
of stellar evolution, neutrinos
 can be produced not as a by-product of thermo-nuclear reactions
({\em thermo-nuclear
 neutrinos}) but directly at expenses of the thermal energy
of stellar matter. According to current physics, {\em cooling}
 neutrinos can originate from several processes, as plasmon
decay (plasma-neutrinos) or photon-electron collisions
(photoneutrinos). Their energy however is lower by about two orders of 
magnitude than the energy of thermo-nuclear neutrinos; their detection
is thus even more difficult than the detection of neutrinos
 from nuclear burning.\\
In this paper, we will discuss all these neutrino sources following the 
 evolution of a suitable sample of stellar structures all along the major 
 phases of H and He burning. The investigation will be completed by
 discussing the relevant contribution given by neutrinos from cooling 
 White Dwarf (WD) structures. In the next section we will discuss
 the properties of thermo-nuclear neutrino sources. Section 3 will be
 devoted to cooling neutrinos both during the nuclear life of a star
 and in the final WD structures. An evaluation of the 
expected galactic and extragalactic neutrino background
 will close the paper.

\section{Thermo-nuclear neutrinos.}

Before entering detailed evolutionary computations, one 
can have a quick look on stellar neutrino sources
by recalling that a fusion of four protons into a He nucleus 
produces 2 $\nu$ and about 27 MeV
of energy which is degraded into eV photons. On this simple
basis, one expects 1 $\nu$ every 10$^7$photons leaving a star. 
One concludes that
stellar neutrinos are closely correlated with
star luminosity. In other words, the neutrino sky should look very
similar to the normal sky of photons, provided that the transparency
of interstellar clouds to neutrinos is taken into
account. Bearing in mind such a
scenario, the production of both `thermo-nuclear' and `cooling' $\nu$
can be evaluated by following the evolution of
a suitable set of stellar models  as chosen to cover the whole
scenario of evolutionary expectations. 

Numerical computations have been performed by adopting the FRANEC
evolutionary code, whose physical inputs have been already reported
in the literature (see, e.g., Castellani et al. 1993, and references therein).
Here we only add that rates for plasma neutrino production are 
from Haft et al. (1994), photoneutrinos are from Itoh et al. (1989), 
neutrinos from pair annihilation processes are from Munakata 
et al. (1985a,1985b), bremstrahlung neutrinos from Dicus et al. (1976),
as corrected by Richardson et al. (1982) and neutrinos produced 
by recombination processes are from Beaudet et al. 1967. 
However, we shall see that in normal stellar evolution 
the bulk of `cooling' neutrino production is given by plasma and photoneutrinos only.

 Evolutionary computations
have been performed over the whole H and He burning phases 
for stars with population I  composition (Y=0.27 Z=0.02 but the model
of 1M$_{\odot}$ for which we adopted Y=0.28 as predicted by the calculations 
of standard solar models) and for selected
values of the stellar mass: M=0.8, 0.9, 1, 3, 5, 7, 9, 16 and 
20 M$_{\odot}$. The choice of the lower limit follows the evidence
that a 0.8 M$_{\odot}$  model spends more than an Hubble time in its central H
burning MS phases. As a consequence, one expects that stars with
 M $\leq 0.8 M_{\odot}$ are still in 
the phase of central H burning, where the luminosity scales with the 
stellar mass according to already known laws (see, e.g., 
Alexander et al. 1997) and the energy production is dominated by
the ppI chain so that only pp neutrinos are produced.

The grid of evolutionary models has been chosen so as to
regularly cover with a sufficient number of different  masses the three 
mass ranges where different evolutionary  behaviours are  expected, namely 
 i) the range of low mass stars 
where stellar cores undergo electron degeneracy during both the H shell
 and the subsequent He shell burning phases, ii) the range of intermediate
mass stars where electronic degeneracy becomes efficient
only in the C,O stellar cores during He shell burning phases, and finally
iii) massive stars where H, He and C are progressively and quietly ignited
in not degenerate stellar cores. 

All models have been followed from the initial Zero Age Main 
Sequence phase up to the carbon ignition or, alternatively,
to the onset of thermal pulses marking the end of the Early
Asymptotic Giant Branch Phase. In this last case, evolutionary computations
have been supplemented with the cooling sequence of a 0.6 M$_{\odot}$
C-O
 white dwarf, assumed to be representative of the final stage of both low and 
intermediate mass stars.
 One has to notice that H and He burning
cover the whole life of low and intermediate mass stars, whereas in massive
stars more advanced nuclear burning phases and further neutrino
production occur before the final
disruption as supernova. However, assuming a 1 M$_{\odot}$ iron core
in the presupernova models, one easily estimates that in these  rapid phases
of evolution the transformation of 1 M$_{\odot}$ of C into Fe
produces a total amount of $\approx$10$^{56}$
 neutrinos, which  can be regarded as a marginal contribution to the 
$\approx$10$^{58}$ neutrinos soon produced by the SN.
Note that the quoted amount of neutrinos represents about 10 percent
of the total amount of neutrinos produced by H burning. This is
what one can predicts on very general grounds, since H burning already
produced He nuclei with the same number of neutrons and protons;
to reach  $^{56}_{26}$Fe two further neutrons over 26 already formed
must be produced, correspondingly accounting for 2/26 of the total
final amount of neutrinos. The actual ratio is moderately larger,
because of the occurrence in stellar matter of a not negligible
amount of original He.

   \begin{figure}[htbp]
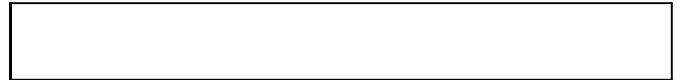

   \picplace{1cm}
      \caption{Neutrino production rate as a function of time
       for  1, 3 and 16 M$_{\odot}$ models.  The vertical line
       in the 1 M$_{\odot}$ model marks the predictions of the 
       Solar Standard Model, as taken at the solar
       age of 4.6 Gyr. 
       }
   \end{figure}

Figure 1 shows the time behaviour of the
neutrino production rate in models with 1, 3 and 16 M$_{\odot}$, chosen as 
representative of the three quoted evolutionary classes. 
Increasing the stellar mass the central temperature
of MS models increases. As a consequence low mass
stars start burning H via the pp chain, whereas in heavier
MS stars
the CNO cycle becomes progressively the dominant mechanism; thus for stars 
with M $\geq$1.5 M$_{\odot}$, CNO neutrinos are dominant
also in the MS phase.
Moreover, as well known, increasing the stellar mass the stellar luminosity
(in photons) increases, driving a corresponding increase of the
neutrino production rate, which in a 22 M$_{\odot}$ MS  model is about 10$^{5.5}$ times
that of a 0.8 M$_{\odot}$.

As a general rule, the evolution off MS  increases both central
temperatures and stellar luminosities. As a consequence, in Fig.1 
one finds that the neutrino production rate tends to increase all along the
H burning phases, whereas CNO neutrinos become progressively more
and more abundant. As already known,  one finds that even 
in low mass stars H burning is eventually dominated by CNO reactions.
After the ignition of central He burning, the only source
of thermo-nuclear neutrinos is the residual efficiency of a H burning shell,
and the neutrino production rate is not more directly correlated to the
photon luminosity of the stars, which is partially powered by
the triple $\alpha$ reactions where no neutrinos are emitted.

   \begin{figure}[htbp]
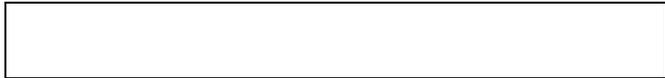

   \picplace{1cm}
      \caption{Thermo-nuclear neutrino yield
 for the models in Fig. 1.}
   \end{figure}

Figure 2 shows the thermo-nuclear neutrino yield
 for the models in Fig. 1 while
 Figure 3 shows the total yield of thermo-nuclear neutrinos for  all
the stars in our sample, until the ignition or at the end of 
He burning, as labeled.
Massive stars have much larger production rate but not 
too much total yield of thermo-nuclear neutrinos. This is because 
massive stars have much shorter lifetimes.  
 In all cases the total number 
of neutrinos (mainly CNO neutrinos) is of the order of 10$^{56}$ - 
10$^{57}$, linearly increasing with the mass (as can be
seen in the lower panel of Fig. 3 the behaviour is linear at least
for masses in the range  3M$_{\odot}<$ M $<20$ M$_{\odot}$).
The total yield of thermo-nuclear neutrinos, as given
by the integral of the neutrino production rate over time,
is directly related to the amount of H converted into He.
The previous result is,
thus, an evidence that the total amount of He produced
 at the end of the nuclear burnings grows linearly
with the mass of the star. 
 One also notes that
at the ignition of He burning the stars
have already produced the total amount of neutrinos from the pp chain
while the CNO neutrinos continue to be produced in the H
 shell active during the He burning phases (upper panel of Fig. 3).


   \begin{figure}[htbp]
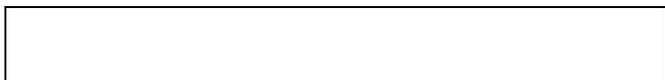

   \picplace{1cm}
      \caption{Upper panel: Total yield of thermo-nuclear
       neutrinos  as a function
      of the stellar mass until the ignition of He burning (dashed line)
	and at the end of the He burning phase (solid line) for the pp
	chain (open circles) and the CNO cycle (filled circles).
	Lower Panel: Total yield of thermo-nuclear neutrinos
       (pp chain + CNO cycle) at the end of He burning.
}
   \end{figure}


\section{Cooling neutrinos}

As an approach to the problem of cooling neutrinos, Figure 4 shows
the evolutionary path of central conditions for selected models in our 
sample as compared with the regions in which the labeled mechanisms
are the dominant for the cooling neutrino production.

   \begin{figure}[htbp]
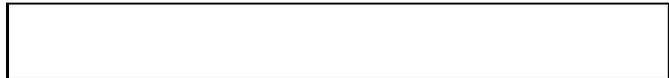

   \picplace{1cm}
      \caption{Evolutionary behaviour of the central temperature and density
for models in our sample. The dashed line
labeled with WD shows the cooling trajectory of a typical 0.6 M$_{\odot}$
White Dwarf. The dominant mechanism for cooling neutrino production
is indicated in each region.
}
   \end{figure}

 It appears that
during their nuclear life, stellar interiors can be mainly affected 
either from photoneutrino or plasma neutrino emissions. 
According to Fig. 4, plasma-neutrinos are important
for low and intermediate mass stars only. For these structures, 
plasma neutrino
cooling is essentially efficient in selected episodes, when  
neutrino production affects the thermal evolution of
degenerate He and/or C+O stellar cores. Figure 5 discloses the 
total yield of photo or plasma 
neutrinos produced until the ignition of He burning  or at the end 
of the He burning phase for the various investigated
masses.

   \begin{figure}[htbp]
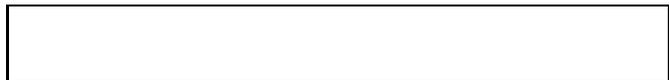

   \picplace{1cm}
      \caption{Total yield of plasma and photo neutrinos
 until the ignition of He burning (upper panel) and at the end of
 He burning phase (lower panel) as a function of the stellar mass.
}
   \end{figure}

As expected, during  H 
burning plasma neutrino cooling is efficient only in less massive
stars, where degenerate He cores develop during the H shell burning
phase. However, over the range of both
low and intermediate mass stars the final yield of plasma-neutrinos
is only weakly dependent on the star mass, with a total 
amount of the order of 5$\cdot10^{55}$ neutrinos.
The reason for such a behaviour is that by increasing the mass, the
time spent in the later phases of He burning where neutrino cooling
attains a renewed efficiency increases, balancing the decreasing production
of neutrinos in the previous H-shell burning phase.

The same figure shows that for 
 M $\geq 3 M_{\odot}$ the total yield of photoneutrinos
can be largely competitive with plasma-neutrinos. 
At the end of H burning the amount
of photoneutrinos dominates over the total yield of plasma-neutrinos 
by several orders of magnitude. During the latest
evolutionary phases of intermediate mass stars 
the plasma-neutrino production rate sensibly increases
and the total yield of plasma and photoneutrinos at the end of He burning
is of the same order of magnitude for stars from 3 to about 9 M$_{\odot}$ 
(lower panel in Fig. 5). For larger masses
the plasma-neutrino production decreases while the photoneutrino
emission remains almost constant at least until 16 M$_{\odot}$. 

%
%

Both low and intermediate mass stars experience a further
episode of neutrino emission during their last phase of cooling as WD,
almost completely due to plasma neutrinos (see fig.4).
 Evolutionary computations have
already disclosed that the cooling neutrino production
affects the cooling of a WD structure just  after the 
exhaustion of nuclear burnings.\\

   \begin{figure}[htbp]
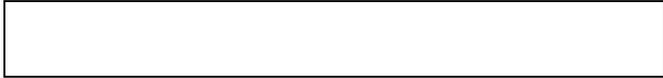

   \picplace{1cm}
      \caption{Time behaviour of the plasma-neutrino
 production rate from a 0.6 M$_{\odot}$ White Dwarf
 (dashed line and right scale) as compared with the
 time behaviour of the stellar luminosity (solid line and left scale).
The big dot shows the initial phase of cooling.}
   \end{figure}

Figure 6 compares the time behaviour of the plasma-neutrino production rate
for a 0.6 M$_{\odot}$ WD, with the time behaviour of (photon) luminosity
showing that
the plasma-neutrino emission is relevant
only in the initial phases of WD, lasting for about 2 Myr and with a
peak emission of $\approx$6$\cdot$10$^{42}$ $\nu$/s. 

As a whole one
expects from each WD a total yield of $\sim$ 3$\cdot10^{56}$
 neutrinos, slightly larger than
 the amount of cooling neutrinos produced in the
previous phases, of the same order of magnitude as
thermo-nuclear neutrinos emitted all along the life of low and intermediate
mass stars. However the energy of plasma/photo neutrino pairs
(20 - 50 KeV) is sensibly lower than that of thermo-nuclear neutrinos
and thus their (future) detection will be correspondigly even more difficult.

Summarizing, stellar evolution gives us some
relevant indications:

i) during their thermo-nuclear phase stars emit 10$^{56}$ - 10$^{57}$ thermo-nuclear 
$\nu$, (almost) linearly depending on the stellar mass,

ii) in all cases the total yield of $\nu$ is dominated by the 
CNO contribution,

iii) during their nuclear life stars emit 3-5$\cdot$10$^{55}$ photoneutrinos,
and a similar number of plasma-neutrinos, these latter only for masses 
$\leq$9 M$_{\odot}$,

iv) The emission of plasma neutrino pairs by cooling
WDs produces about 10$^{56}$ neutrinos, i.e., a number of the
same order of magnitude  
of the number  of cooling neutrinos in previous nuclear 
phases.

\section{Galactic neutrinos}

Adopting from Cox \& Mezger (1989) a galactic disk luminosity
of 4.5$\cdot$10$^{10}$ L$_{\odot}$, the
thermo-nuclear neutrino
production rate is easily found as  
8$\cdot$10$^{48}$ s$^{-1}$. To obtain more detailed predictions,
one needs to estimate the neutrino fluxes from stars of various ages
and chemical composition. As expected on very general grounds, 
and confirmed by numerical experiments
 (Raimondo 1995), assumptions about the original composition of the
stellar structures play  a minor role in our problem.
Passing from a solar composition to 
extreme metal poor populations (Z=0.0001), for each given mass the total 
amount of neutrinos emitted by a star keeps constant within, about, 10\%.
Differences in the various types of neutrinos can reach a factor 2 or 3,
depending on the stellar mass. To reach an order of magnitude estimate
we will neglect the effects of chemical composition, interpolating the 
evolutionary results discussed in the previous sections
to determine a theoretical value of both
thermo-nuclear and cooling neutrino production rates for each value of the
stellar mass and for each assumption about the age of the
structure. A similar data base can thus be used to estimate the
contribution of stars in our galaxy to the flux of neutrinos
expected on earth.  

However, to predict the neutrino spectrum
one needs a model for the galactic stellar population.
For this purpose we followed the galactic model of
Bahcall (1986). We distributed 
6$\cdot$10$^{10}$ M$_{\odot}$ of stars in the galactic disc, 1$\cdot$10$^{10}$
 M$_{\odot}$
 in the 
galactic bulge, neglecting the contribution from the
1-3$\cdot$10$^{9}$ M$_{\odot}$ expected in the galactic halo. Again, we remark
 that 
these figures are to be taken as order-of-magnitude estimates.
As a matter of the fact, the `what you see is what you get' model
by Fich \& Tremaine (1991) gives for the disk a larger
mass; M$_{disk}\simeq 10^{11}$ M$_{\odot}$, with only a minor
contribution from interstellar matter.
Interesting enough, one finds that even these order of magnitude
estimates give constraints on the history of Star Formation Rate (SFR):
 adopting, e.g.,
a SFR constant with time one cannot reproduce the galactic
M/L ratio for any reasonable assumption about the Initial Mass Function (IMF).

 \begin{table}[h]
  \begin{center}
  \caption{Estimated production rate of thermo-nuclear neutrinos and
of cooling neutrino-antineutrino pairs from the Galaxy}
\label{tab:flussi1}

  \small

  \begin{tabular}{c c c c}
\hline
\hline
  {} &{} &{} &{} \\
reactions      & Disk  &  Bulge  & Total\\
 s$^{-1}$ &{} &{} \\
\hline

  {} &{} &{} &{}\\

 $\nu_{pp}$     &  1.4 $10^{48}$   &  1.9 $10^{47}$   & 1.6 $10^{48}$ \\ 

 $\nu_{7Be}$    &  1.4  $10^{47}$   &  1.8  $10^{46}$  & 1.5  $10^{47}$\\ 

 $\nu_{8B}$     &  5.7  $10^{45}$   & 4.0  $10^{44}$   & 6.1  $10^{45}$\\ 
\hline
 $\nu_{13N}$    &  2.7  $10^{48}$  & 2.9  $10^{47}$   & 3.0  $10^{48}$\\ 

 $\nu_{16O}$    &  2.7  $10^{48}$  &  2.9  $10^{47}$  & 3.0  $10^{48}$\\ 
 \hline
 plasma (stars) &   2.5  $10^{48}$   & 2.1  $10^{47}$ &  2.7  $10^{48}$ \\ 
 plasma (WD)    &   4.8  $10^{48}$   & 5.8  $10^{47}$ &  5.4  $10^{48}$  \\
\hline
photo            &  4.0  $10^{46}$   &      0         & 4.0  $10^{46}$      \\ 
  {} &{} &{}\\
\hline
\hline

  \end{tabular}
  \end{center} 
  \end{table}

According to indications given in the recent literature (see, e.g., De Marchi 
\& Paresce 1997) we adopted a Salpeter IMF for masses
 larger than 0.2 M$_{\odot}$ and a
flat distribution below this limit, down to 0.1 M$_{\odot}$.
With this choice,
the galactic M/L is well reproduced by a disk with SFR  of the form
exp(-t/$\tau$) with $\tau$ of the order of 2.5 Gyr. For the galactic bulge,
a single episode of star formations 10$^{10}$ years 
ago automatically accounts for the bulge luminosity (L$_{bulge}\simeq
5 \cdot 10^{9}$ L$_{\odot}$) as  estimated by Cox \& Mezger (1989). 
To evaluate the galactic neutrino production rate we used a
Monte Carlo technique, randomly disseminating stars with masses
and ages constrained by the above prescriptions, adding the contribution of 
the single sources to the final net galactic neutrino production rate.
Table 1 gives the total neutrino production rate by the Galaxy
according to the above quoted assumptions.
It appears that in both disk and bulge populations the neutrino
production is dominated by CNO neutrinos, as expected since the total
photon luminosity is dominated by the most luminous, CNO burning stars.
Moreover the balance among the
various nuclear sources of
neutrinos appear not too much dependent from the age of the stellar
population.


 \begin{table}[h]
  \begin{center}
  \caption{Estimated neutrino fluxes at the earth surface.
}
\label{tab:flussi2}

  \small

  \begin{tabular}{c c c c}
\hline
\hline
  {} &{} &{} &{} \\
 Flux   & Disk  &  Bulge  & Total\\
 cm$^{-2}$ s$^{-1}$ &{} &{} \\
\hline

  {} &{} &{} &{}\\

 $\nu_{pp}$     &  420      &  22   & 442 \\ 

 $\nu_{7Be}$    &   42      &   2   &  44 \\ 

 $\nu_{8B}$     &   1.8     & 0.05  &   2 \\ 

Total pp         &         &        &  488   \\
\hline
 $\nu_{13N}$    &   810      &  33   &  843  \\ 

 $\nu_{16O}$    &   810      &  33   &  843  \\ 

Total CNO       &            &        & 1686 \\
 \hline
 plasma (stars) &   750      &   24   &  774  \\ 
 plasma (WD)    &   1440     &   67   & 1507  \\
\hline
photo           &    12      &     0  &   12   \\ 
  {} &{} &{}\\
\hline
\hline
  \end{tabular}
  \end{center} 
  \end{table}

Table 2 gives the neutrino fluxes on earth showing the contribution of
the disk and of the bulge.   
As a whole, the Galactic neutrino flux is $\approx2\cdot10^3$ 
cm$^{-2}$ s$^{-1}$ thermo-nuclear neutrinos
which appears in agreement with the predictions
by Hartmann et al. (1994) quoted in the introduction of this paper. 
A similar figure is found for cooling neutrinos.
Figure 7 shows the energy spectrum of galactic thermonuclear neutrinos
compared with the theoretical solar neutrino spectrum from Bahcall \& Ulrich 1988.

   \begin{figure}[htbp]
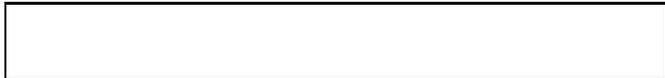

   \picplace{1cm}
 \caption{ Galactic neutrino spectrum (continuos line) compared with the solar neutrino
spectrum from Bahcall \& Ulrich 1988 (dashed line). The neutrino fluxes from continuos
sources are given in the unit of number per cm$^2$ per second per MeV and the line fluxes 
are given in number per  cm$^2$ per second.
}
   \end{figure}

It is interesting to  notice that present estimates of the galactic 
neutrino background appear in fair agreement with the corresponding
background of photons.
As quoted in the introduction 
of this paper, one expects neutrino and photon fluxes to be 
fairly stringently correlated, with 2 expected neutrinos every
26.7 MeV in photons (clearly this is true only in the approximation
in which one neglects
the energy arising from He burning which is expected to be
a minor fraction of the total luminosity emitted by the Galaxy). 
For the galactic background Mathis et al. (1983)
give the photon flux on earth in the wavelenght range 0.09-8 $\mu$m,
due only to stars: 1.3$\cdot$10$^{10}$ eV s$^{-1}$ cm$^{-2}$,
 which would  imply
a neutrino flux of about 950 $\nu$ cm$^{-2}$ s$^{-1}$.
in agreement with present theoretical expections. 

   \begin{figure}[htbp]
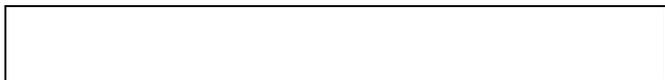

   \picplace{1cm}
 \caption{The angular distribution of the total neutrino flux for the present
galactic model (see text) in terms of the galactic longitude $\phi$ 
(the center of the Galaxy
is at $\phi$=0).
}
   \end{figure}
Figure 8 shows the angular
distribution of the total neutrino flux in terms of the galactic
longitude, $\phi$. One can distinguish three components:\\
i) an isotropic component, due to the nearest stars, of about
 3 $\nu$ cm$^{-2}$s$^{-1}$degree$^{-1}$.
This contribution is reduced to
$\approx$1 $\nu$ cm$^{-2}$s$^{-1}$degree$^{-1}$
 if we include in the computation
only stars distant from the Sun more than 1 Kpc.\\
ii) the contribution of the bulge which decreases rapidly to zero 
in few degrees away from the direction of the galactic center.\\
iii) the disk component, which, in agreement to the adopted
star distribution, decreases to about 1/3 for a galactic longitude of
about 50$^o$. \\
One notes that stars within 1 Kpc$^{3}$ from the
Sun contribute by about 1/3 to the total neutrino flux.

\section{Cosmic background}

From the estimate of the neutrino
output of our Galaxy one easily obtains a corresponding
estimate of the stellar contribution to the cosmic background of neutrinos
if, on the average, galaxies radiated 
over all the time as our own Galaxy does at present time.
As a matter of fact, the density of neutrinos can be simply obtained 
by multiplying the neutrino production rate per
galaxy by the spatial density of galaxies in the Universe 
and by the time elapsed since the appearance of galaxies.
The flux is finally given
by the neutrinos density times the speed of light.
We already found that our Galaxy produces about 8$\cdot$10$^{48}$ $\nu$/s.
 Since, on the average,
galaxies are slightly more massive than our own
 Galaxy, let us scale this figure to 
a mean galaxy with about 10$^{11}$ M$_{\odot}$, thus adopting a 
neutrino production rate
of about 1$\cdot$10$^{49}$ $\nu$/s. 
For such a rough estimate of the cosmic background one can multiply the galactic
neutrino production rate  for the mean density of galaxies (0.01 Mpc$^{-3}$, see e.g. 
Peebles 1971) and for a reference  galactic age of 10 Gyr.  
For thermo-nuclear neutrinos, the result is a background density 
of 10$^{-9}$ $\nu$ cm$^{-3}$ and a flux of 40 $\nu$ cm$^{-2}$ s$^{-1}$.
A corresponding figure is for cooling neutrinos.
We already noted that  the balance among the
various nuclear sources of
neutrinos is not dramaticaly dependent from the stellar
population, thus data in previous Tables 1 and 2 could also be used
to give a rough estimate of the relative abundances of thermo-nuclear neutrinos and
the corresponding  energy spectrum.\\
One may finally notice that the comparison between 
observational values and theoretical
predictions is not a safe procedure for the cosmic background 
due to the presence of
sources other than nuclear which contribute to this quantity (e.g. the
 active galactic nuclei). However, one can adopt the present theoretical
result as a rough estimate of the stellar contribution to the cosmic
background. If one takes into account that galaxies are suggested
 to have been more
 luminous in the past, the present result can be regarded
 as a lower limit for the contribution to the cosmic
 neutrino background given by stars during their `normal' thermo-nuclear
life and their final cooling phase.\\
As a whole, one finds that the above estimates are only marginally
affected by the adopted model of the galaxy, the key parameter
remaining the assumed galactic luminosity. In particular, reasonable 
variations of the adopted IMF affects the galactic M/L and/or the SFR
but with negligible influence on the given order of magnitudes
for neutrinos backgrounds.   
For the sake of comparison, one may recall that the additional contribution
by supernovae to the cosmic neutrino background has been estimated
of the order of $\approx$1 $\nu$ cm$^{-2}$ s$^{-1}$ (Woosley et al. 1986,
Hartmann et al. 1994). A figure, however, to be regarded as a lower limit,
since several arguments suggest a possible increase of the fluxes up to one
or two orders of magnitudes (see Krauss et al. 1984, Woosley et al. 1986).
In this context, one may notice that if a SN emits 3$\cdot10^{53}$ erg
 under the
form of neutrinos with mean energy 13 MeV (Woosley et al. 1986), the total
neutrino yield is given by 1$\cdot10^{58}$ $\nu$. It follows that in our Galaxy
2 SN per century would give the same yield of neutrinos given by
all the quiescent stars in the Galaxy.

\section{Final remarks}

We presented detailed evolutionary computations
for estimating the total yield of neutrinos (both from nuclear and
cooling origin) produced by stars of various masses all 
along their nuclear life and the final cooling as WD.
Adopting a reasonable modelization of the Galaxy we predicted
the flux of stellar neutrinos at the earth surface, giving
also a lower limit for the flux of cosmic neutrinos. In all cases,
we find that CNO neutrinos dominate the flux, with a
distribution among the various nuclear  sources
which appears not considerably dependent on the assumed stellar
population. 

Before closing the paper,
it may be worth noticing that the adopted galactic M/L ratio, as 
taken from the current literature, does not support  
the often quoted nuclear evidences for a larger luminosity of galaxies in
the past, as far as our own Galaxy is concerned.
According to an argument early presented by Hoyle \& Tayler (1964),
and recently reviewed by Reeves (1994), if the energy 
presently emitted by a ``mean'' spiral galaxy
is evaluated of the order of 2$\cdot$10$^{-13}$ eV/nucleon s$^{-1}$,
 and if this energy output
remained constant over the time, thus one finds that
all along the life of galaxies, only 1.4 \% of H can have
been converted in He, against the observational evidences which suggest
an increase of the abundance by mass of He of the order of 5\%.
This led to the hypothesis that galaxies were more luminous in the past.

However, adopting for the Galaxy, as we did, L$\simeq$ 4$\cdot$10$^{10}$
L$_{\odot}$ and M$\simeq$ 7$\cdot$10$^{10}$ M$_{\odot}$, one would derive
 an energy output 
per nucleon about 6 times larger, or 4 times larger if one assumes a galactic
mass of 1$\cdot$10$^{11}$ M$_{\odot}$.
 Accordingly, one finds that for a constant 
luminosity 5-8 \% of H is allowed to have been converted in He, 
in agreement with observational constraints. 
We do not claim that the  galactic luminosity was constant in the
time: we only drive the attention on this matter to stimulate further
discussion on a point that is well beyond the purpose
of this paper.


\begin{thebibliography}{9}


\bibitem{alex} Alexander D.R., Brocato E., Cassisi S., Castellani V., Ciacio F.,
Degl'Innocenti S., 1997, A\&A 317, 90 
\bibitem{BahSon} Bahcall J. N., 1986, ARA\&A 24, 577 
 \bibitem{Bah} Bahcall J. N., 1989, `` Neutrino Astrophysics'', Cambridge
      University Press, Cambridge, England
\bibitem{} Bahcall J.N., Ulrich R.K., 1988, Rev. Mod. Phys. 60, 297
\bibitem{} Bahcall J.N., Davis R. Jr., Parker P., Smirnov A., Urlich R.
 (editors), 1995, ``Solar Neutrinos the first thirty years'' , Addison-Wesley
publishing Company
  \bibitem{Beau}Beaudet G., Petrosian V., Salpeter E.E., 1967, ApJ 150,979
  \bibitem{Cast}Castellani V., Degl'Innocenti S., Fiorentini G., 1993,
      A\&A 271, 601.
  \bibitem{Coxm}Cox P., Mezger P.G., 1989, AAR 1, 49
\bibitem{} De Marchi G., Paresce F., 1997, ApJ 476, L19
  \bibitem{Dicu}Dicus D.A., Kolb E.D., Schramm D.N., Tubbs D., 1976,
 ApJ 210, 481
  \bibitem{Fich}Fich M., Tremaine S., 1991, ARA\&A 29, 409
  \bibitem{Haft}Haft M., Raffelt G., Weiss A., 1994, ApJ 425, 222.
  \bibitem{Hart}Hartman D., Meyer B., Clayton D., Luo N., Krishnan T.,
 1994, in "Nuclei in the Cosmos III", AIP conference proceedings n.327,
 ed. M.Busso, R.Gallino, C.M.Raiteri p.447.
\bibitem{} Hoyle F., Tayler R.J., 1964, Nature 203, 1108
\bibitem{Itoh}Itoh N., Adachi T., Nakagawa M., Kohyama Y., 1989,
  ApJ. 339, 354. 
\bibitem{Krauss} Krauss L. M., Sheldon L. G., Schramm D. N., 1984, Nature 310, 191 
\bibitem{Mathi} Mathis J.S., Mezger P.G., Panagia P., 1983, A\&A 128, 212 
 \bibitem{Muna} Munakata H., Kohyama Y., Itoh N., 1985a, ApJ 296,197 
\bibitem{Muna2} Munakata H., Kohyama Y., Itoh N., 1985b, ApJ 304, 508  
\bibitem{} Peebles P.J.E., 1971, in ``Physical Cosmology'', ed. A. Wightmann and J.
Hopfield, Princeton University Press, Princeton, New Jersey.
\bibitem{Gabri} Raimondo G., 1995, Diploma thesis,
  University of Pisa
\bibitem{} Richardson M.B., Van Horn H.M., Ratcliff K.F., Malone R.C., 1982, ApJ 255, 624
\bibitem {} Reeves H., 1994, Rev. Mod. Phys. 66, 193
\bibitem {} Woosley S.E., Wilson J.R., Mayle R., 1986, ApJ 302, 19
 \end{thebibliography}
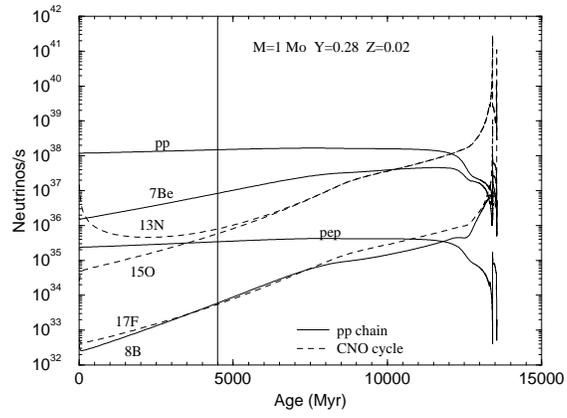
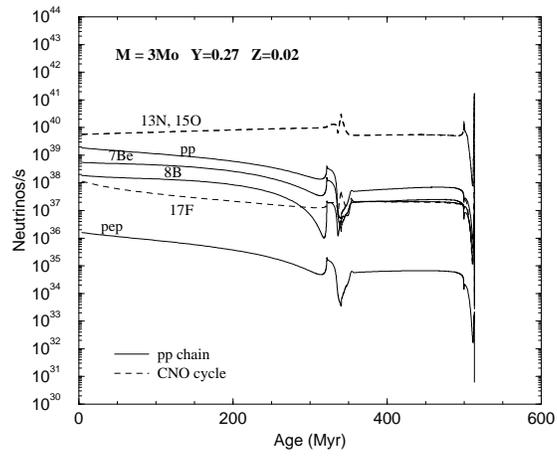
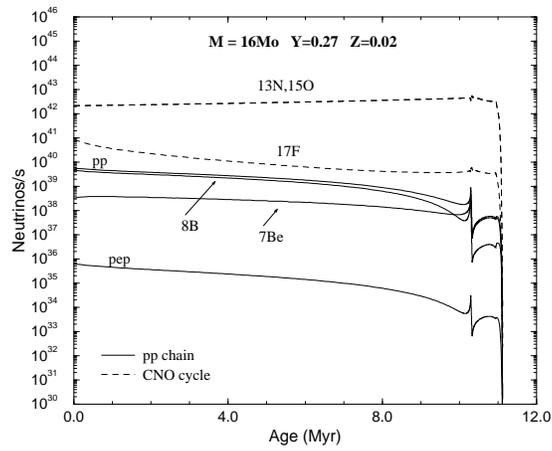
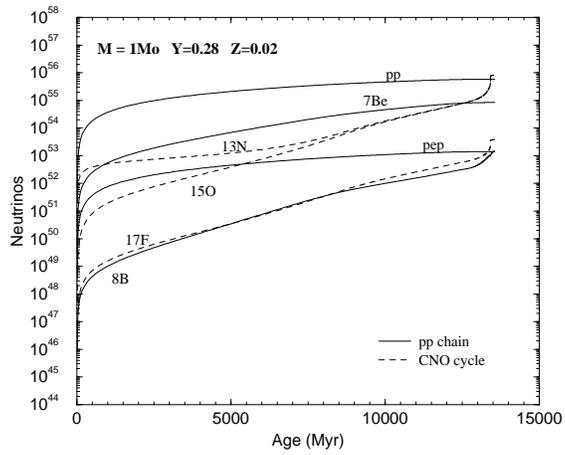
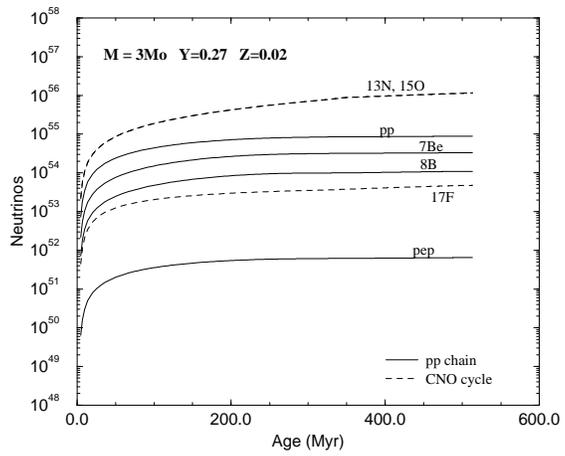
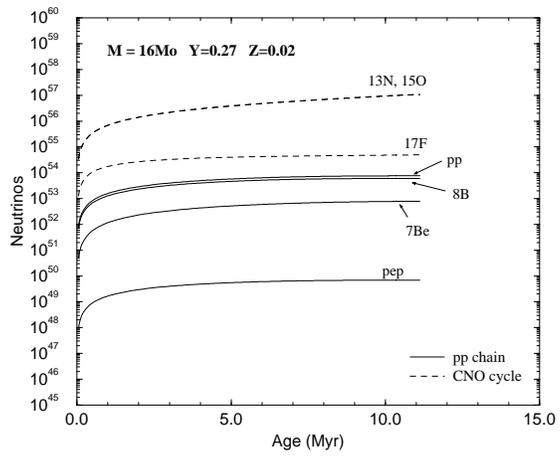
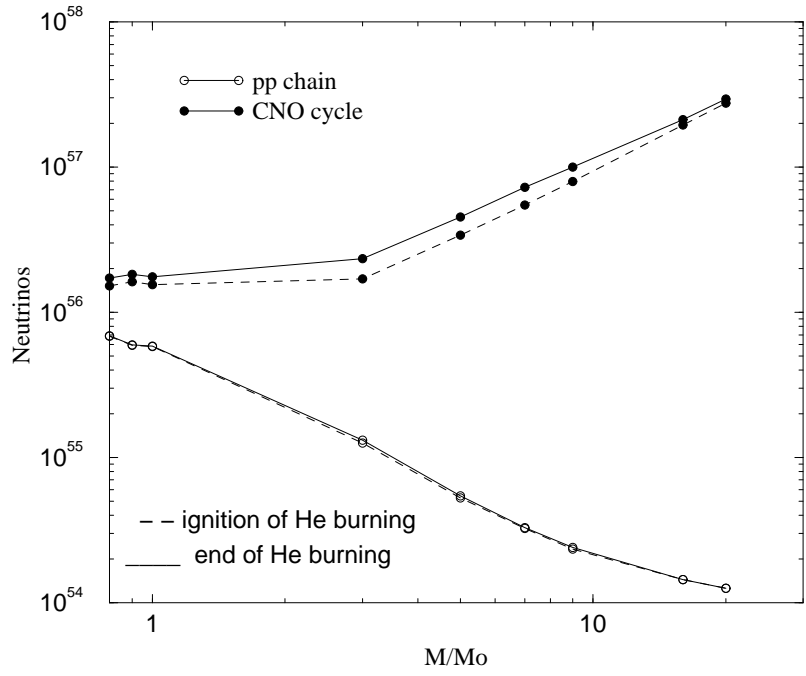
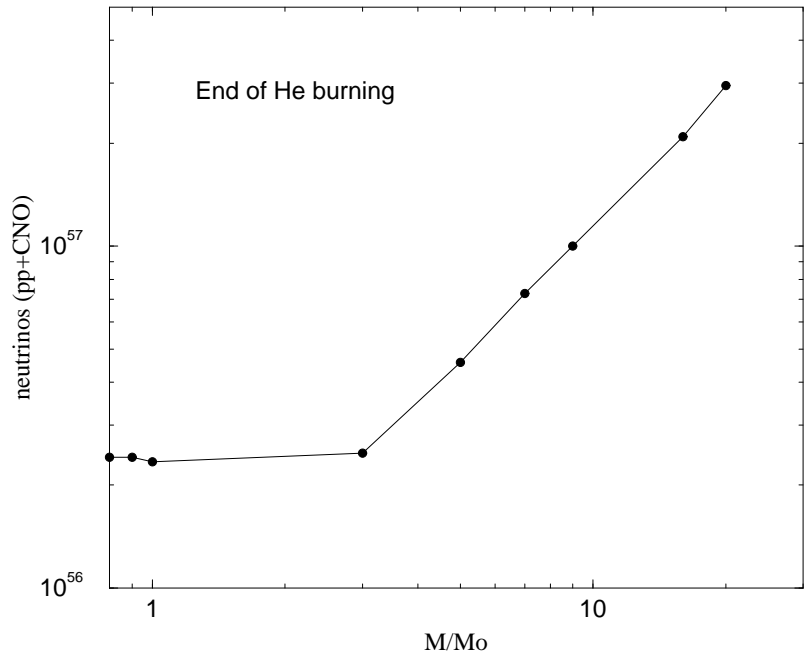
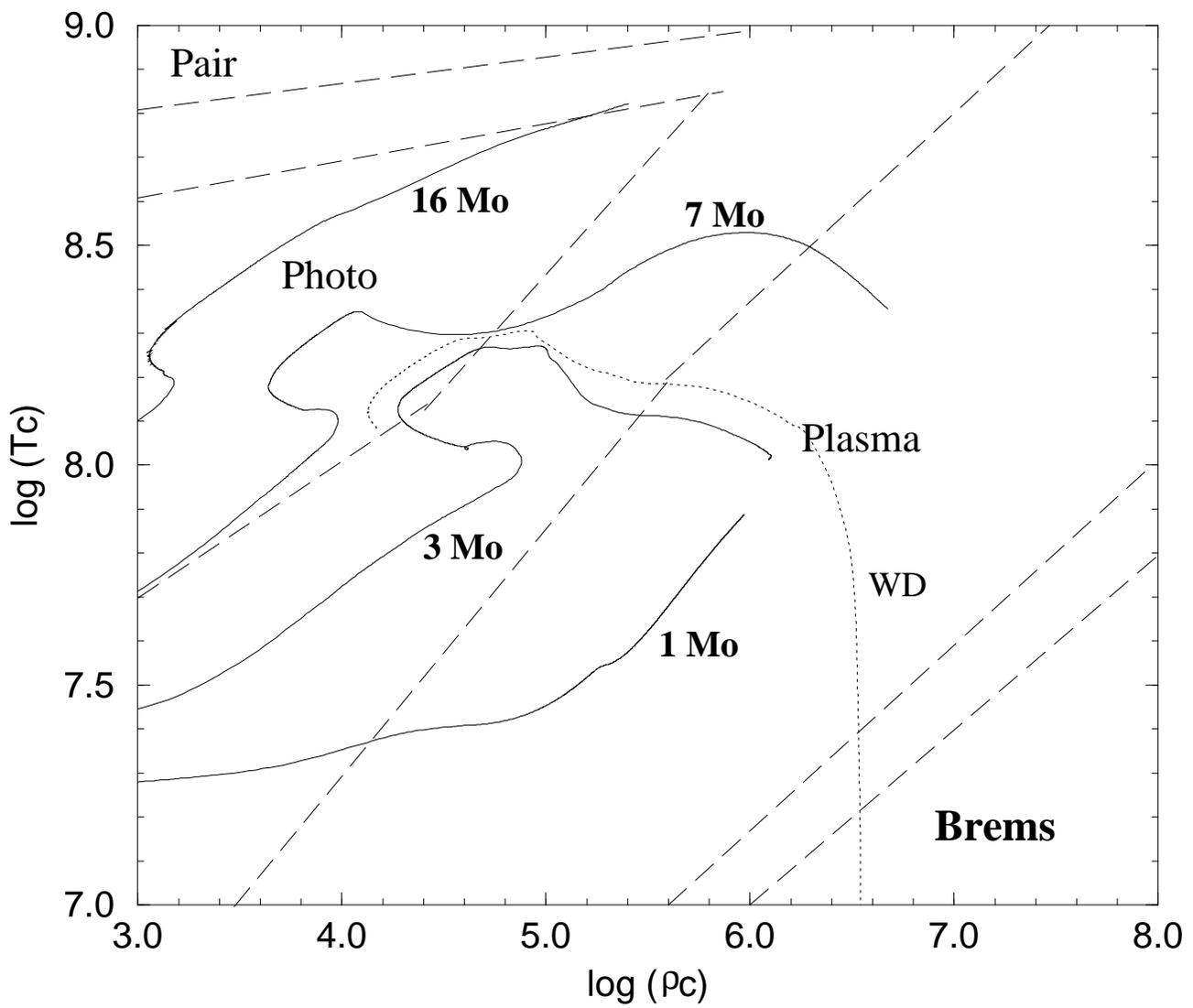
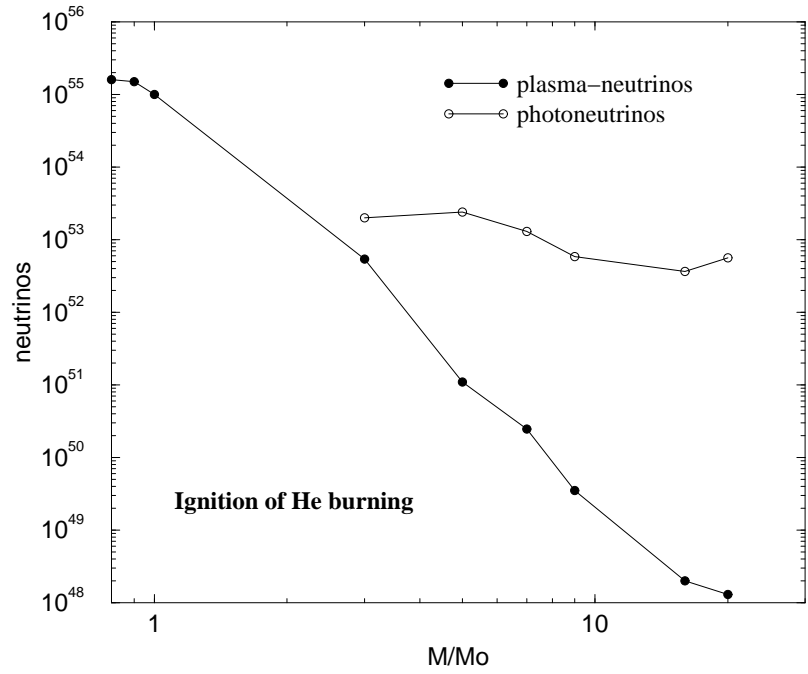
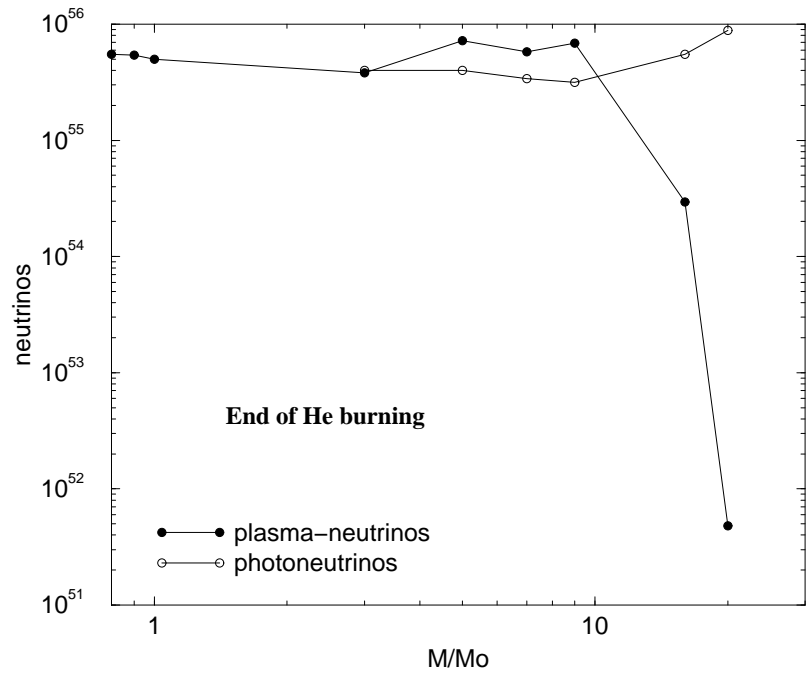
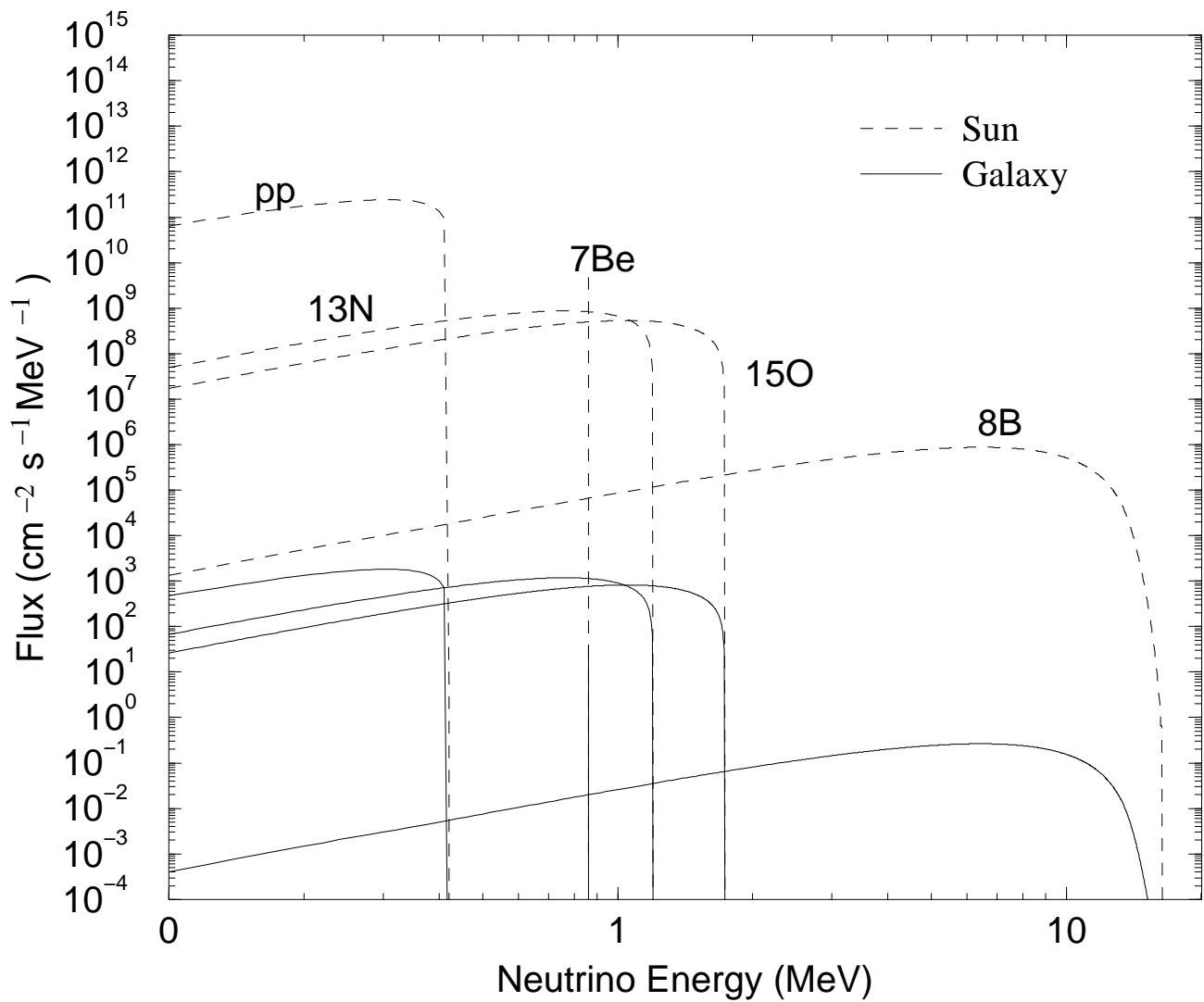
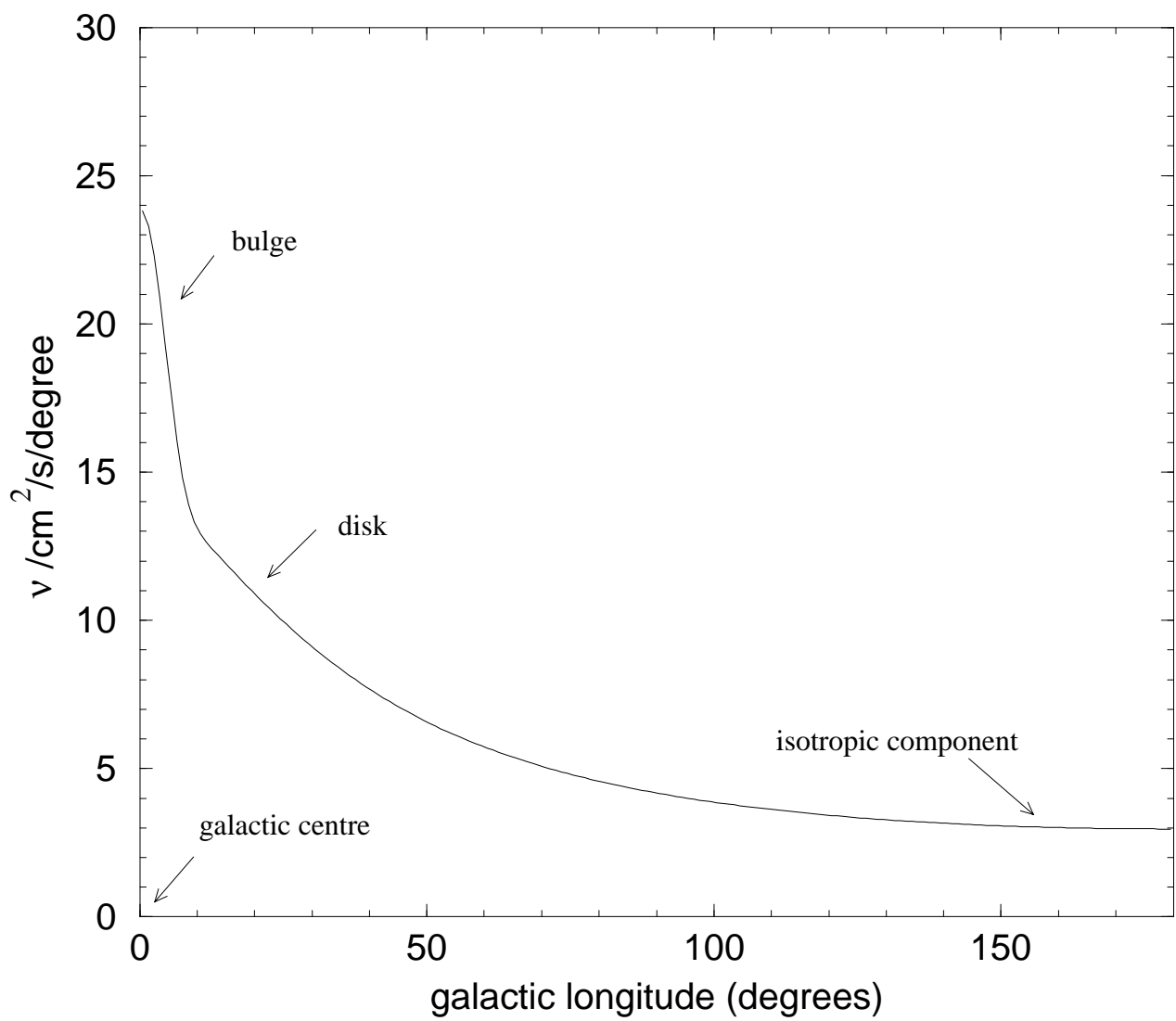
\end{document}